\documentclass[aps,prb,amsmath,amssymb,footinbib,showpacs,twocolumn,superscriptaddress]{revtex4-2}
\usepackage{amsmath}
\usepackage{amssymb}
\usepackage{amsthm}
\usepackage{setspace}
\usepackage{graphicx}
\usepackage{braket}
\usepackage{mathrsfs}
\usepackage{float}
\usepackage[colorlinks = true,linkcolor = blue,urlcolor  = blue,citecolor = blue,anchorcolor = blue]{hyperref}
\usepackage[utf8]{inputenc}
\usepackage[english]{babel}
\usepackage{bm}
\usepackage{csquotes}

\begin{document}

\title{Magnetotransport in spin-orbit coupled noncentrosymmetric and Weyl metals}
\author{Gautham Varma K}
\affiliation{School of Physical Sciences, Indian Institute of Technology Mandi, Mandi 175005, India.}
\author{Azaz Ahmad}
\affiliation{School of Physical Sciences, Indian Institute of Technology Mandi, Mandi 175005, India.}
\author{Sumanta Tewari}
\affiliation{Department of Physics and Astronomy, Clemson University, Clemson, SC 29631, USA.}
\author{G. Sharma}
\affiliation{School of Physical Sciences, Indian Institute of Technology Mandi, Mandi 175005, India.}

\begin{abstract}
Recently, chiral anomaly (CA) has been proposed to occur in spin-orbit coupled non-centrosymmetric metals (SOC-NCMs), motivating CA to be a Fermi surface property rather than a Weyl node property. Although the nature of the anomaly is similar in both SOC-NCMs and Weyl systems, here we point out significant fundamental differences between the two. We show that the different nature of the orbital magnetic moment (OMM) in the two systems leads to nontrivial consequences-- particularly the sign of the longitudinal magnetoconductance always remains positive in a SOC non-centrosymmetric metal, unlike a Weyl metal that displays either sign. Furthermore, we investigate the planar Hall effect and the geometrical contribution to the Hall effect in the two systems and point out significant differences in the two systems. We conduct our analysis for magnetic and non-magnetic impurities, making our study important in light of current and upcoming experiments in both SOC-NCMs and Weyl metals.
\end{abstract}

\maketitle

\section{Introduction} Chiral anomaly roots its origin in high-energy physics~\cite{adler1969axial,bell1969pcac}. It refers to the non-conservation of left and right-handed Weyl fermions separately in the presence of external gauge fields. Over the past decade, its unexpected appearance in solid-state systems has caused great excitement in the condensed matter community~\cite{armitage2018weyl,volovik2003universe,nielsen1981no,nielsen1983adler,wan2011topological,xu2011chern,zyuzin2012weyl,son2013chiral,kim2014boltzmann,lundgren2014thermoelectric,cortijo2016linear,sharma2016nernst,zyuzin2017magnetotransport,nandy2017chiral,das2019berry,kundu2020magnetotransport,knoll2020negative,sharma2020sign,bednik2020magnetotransport,he2014quantum,liang2015ultrahigh,zhang2016signatures,li2016chiral,xiong2015evidence,hirschberger2016chiral,son2012berry,goswami2013axionic,goswami2015axial,zhong2015optical}. Specifically, Weyl fermions, discovered as electronic excitations in specific systems (termed as Weyl semimetals (WSMs)), can manifest the anomaly that can be detected via relatively simple transport~\cite{son2013chiral,kim2014boltzmann,lundgren2014thermoelectric,cortijo2016linear,sharma2016nernst,zyuzin2017magnetotransport,nandy2017chiral} or optical~\cite{goswami2015optical, levy2020optical, parent2020magneto, levy2020optical,song2016detecting,rinkel2017signatures,yuan2020discovery,cheng2019probing} measurements. The key requirement is that the elementary excitations should be chiral and relativistic in odd spatial dimensions~\cite{fujikawa2004path}. 

The realization of the anomaly has recently been extended to certain other systems distinct from Weyl semimetals~\cite{gao2017intrinsic,dai2017negative,andreev2018longitudinal,wang2018intrinsic,nandy2018berry,fu2020quantum,pal2021berry,wang2021helical,PhysRevB.105.L180303,das2023chiral}. Specifically, it has been suggested that the anomaly can be realized in spin-orbit-coupled (SOC) non-centrosymmetric metals (NCMs) that host nonrelativistic fermions with only one relevant band touching point~\cite{PhysRevB.105.L180303}. The effect of the anomaly on charge and thermal transport properties of SOC-NCMs has recently been studied, and it has been suggested that the anomaly results in positive longitudinal magnetoconductance (LMC)~\cite{PhysRevB.105.L180303,das2023chiral}, akin to Weyl semimetals. The sign of LMC has been a subject of much debate and exploration in WSMs. It is expected to crucially depend on the nature of impurities, the strength of the magnetic field, and the strength of the intervalley scattering. Under strong magnetic fields, due to Landau quantization, the LMC sign depends on the nature of scattering impurities~\cite{goswami2015axial,lu2015high,chen2016positive,zhang2016linear,shao2019magneto,li2016weyl,ji2018effect}. 

\begin{figure*}
    \centering
    \includegraphics[width=1.99\columnwidth]{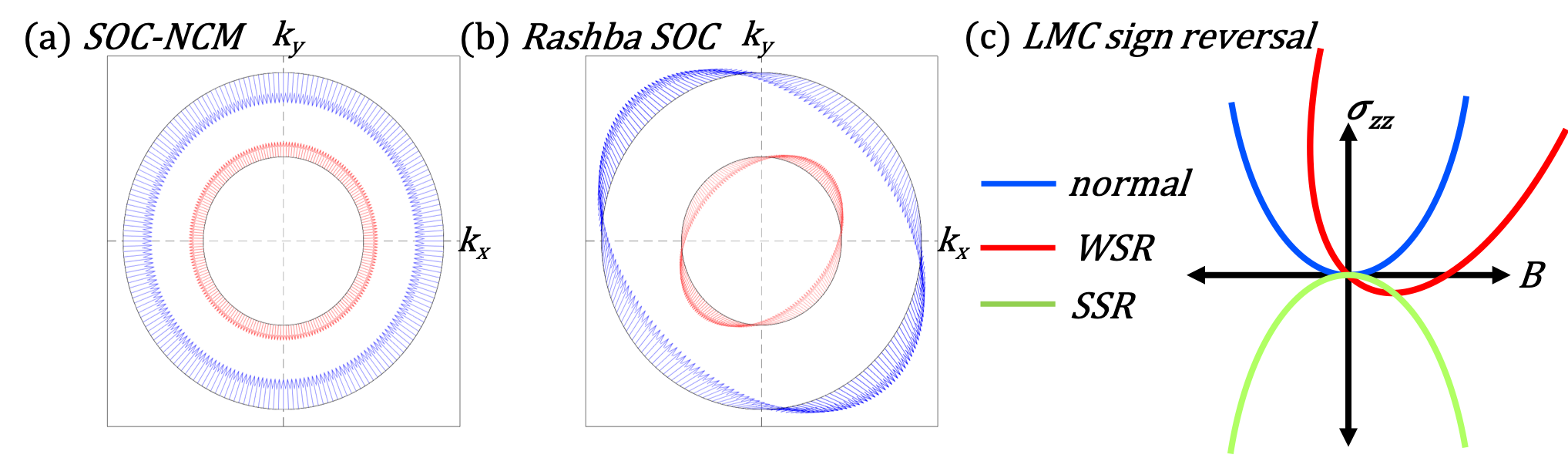}
    \caption{Spin texture of (a) SOC-NCM in the $k_z=0$ plane, and (b) a Rashba coupled system. The arrows point in the direction of the spin expectation value $\mathbf{S}^\lambda$. Blue and red arrows are for the outer and inner Fermi surfaces, respectively. (c) Schematic figure representing weak-sign-reversal (WSR) and strong-sign-reversal (SSR) compared to normal LMC in Weyl systems.}
    \label{fig:spin_text}
\end{figure*}

Recently, we pointed out that the sign of LMC is, in fact, more nuanced~\cite{ahmad2023longitudinal}. LMC in Weyl systems can typically be expressed as $\sigma_{zz} = \sigma_0 + \sigma_{zz}^{(2)} (B-B_0)^2$. `Strong-sign-reversal' is characterized by the reversal of orientation of the magnetoconductance
parabola with respect to the magnetic field, while in `weak-sign-reversal,' the magnetoconductivity depends on the direction of the magnetic field and 
is not correlated with the orientation of the LMC parabola. 
Fig.~\ref{fig:spin_text} (c) shows a schematic description of strong and weak-sign-reversal of LMC. Specifically, in the case of weak-sign-reversal, LMC is linear near zero magnetic field, while the vertex of the parabola ($B_0$) is shifted from the origin, but the quadratic coefficient of LMC ($\sigma_{zz}^{(2)}$) remains positive. In the case of strong-sign-reversal, importantly, the quadratic coefficient $\sigma_{zz}^{(2)}$ becomes negative. 
When Landau quantization can be ignored under weak magnetic fields, quasiclassical Boltzmann analysis suggests that sufficiently strong intervalley scattering can reverse the sign of LMC from positive to negative (strong-sign-reversal)~\cite{knoll2020negative,xiao2020linear,sharma2020sign}. Whether or not the longitudinal magnetoconductance in SOC-NCMs shows similar characteristics also remains an important and pertinent question in the field. Furthermore, the focus of all the previous works has been particularly on point-like scalar non-magnetic impurities. The fate of LMC in both spin-orbit-coupled and Weyl metals in the presence of (pseudo)magnetic impurities remains to be determined. 

\begin{figure}
    \centering
    \includegraphics[width=0.98\columnwidth]{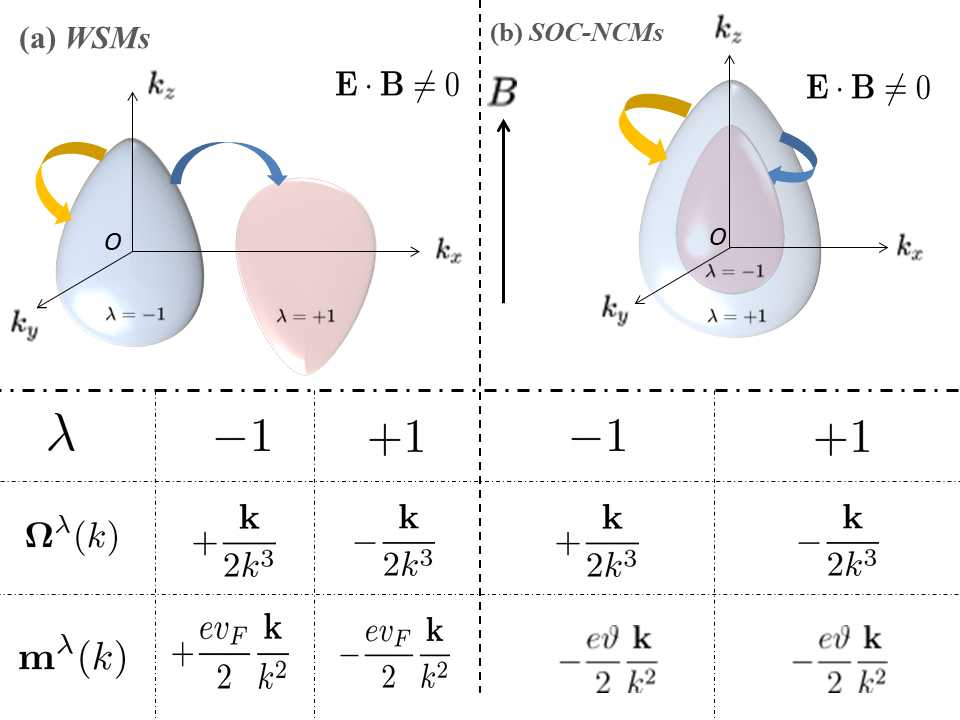}
     \caption{Quasiparticle scattering in WSMs (a) and SOC-NCMs (b). Unlike WSMs, quasiparticle scattering in SOC-NCMs occurs between the surfaces (FSs) associated with a single nodal point. The two Fermi FSs in SOC-NCMs have opposite Berry curvature ($\boldsymbol{\Omega}^\lambda(\mathbf{k})$), but crucially, unlike WSMs, have the same orbital magnetic moment ($\mathbf{m}^\lambda(\mathbf{k})$). Blue and yellow arrows represent the internode (interband for SOC-NCMs) and intranode (intraband for SOC-NCMs) scattering, respectively, in WSMs. Here, $\lambda$ is the band/node index. The oval shape of the Fermi surfaces is due to the coupling of the orbital magnetic moment in an external magnetic field.}
\label{fig:Kramer_weyl_Node_f}
\end{figure}

In SOC-NCMs, we focus on the vicinity of one nodal point surrounded by two Fermi surfaces as depicted in Fig.~\ref{fig:Kramer_weyl_Node_f} (b). This is in contrast to the two separate nodal points and Fermi surfaces we are concerned with in WSMs (Fig.~\ref{fig:Kramer_weyl_Node_f} (a)). The role of intranode scattering in WSMs is replaced by intraband scattering in SOC-NCMs. This scattering preserves the chirality of the scattered quasiparticles. Internode scattering in WSMs is equivalent to interband scattering in SOC-NCMs, reversing the quasiparticle chirality. Internode scattering in WSMs requires the transfer of large momentum of the order of separation between the Weyl nodes, which is usually weaker than intranode scattering requiring a small momentum transfer. In contrast, in SOC-NCMs, the momentum transfer with interband scattering is not necessarily small, as both the Fermi surfaces surround a single nodal point. Thus, interband scattering is expected to be at least as significant in SOC-NCMs as it is in WSMs, and its exploration remains an open question.

In this {work}, we probe the role of interband scattering in SOC-WSMs and show that in the quasiclassical low-field regime, unlike WSMs, the sign of LMC is \textit{not} sensitive to the relative strength of the interband scattering. Longitudinal magnetoconductance in SOC-NCMs is found to be always positive, irrespective of the strength of the interband scattering. We trace the reason to the orbital magnetic moment (OMM) in SOC-NCMs that is of equal magnitude and sign at both the bands, as compared to the case of WSMs where OMM has equal magnitudes but opposite signs at the two nodes (see Fig.~\ref{fig:Kramer_weyl_Node_f}). We examine how the subtle difference in OMM can lead to drastic differences in other transport properties, such as the planar Hall conductivity, and also give rise to a finite geometrical contribution to the Hall conductivity in SOC-NCMs. Furthermore, we also analyze all the properties of both WSMs and SOC-NCMs in the presence of both point-like scalar and magnetic impurities, which has remained an open problem so far. 

 \begin{figure}
    \centering
    \includegraphics[width=.98\columnwidth]{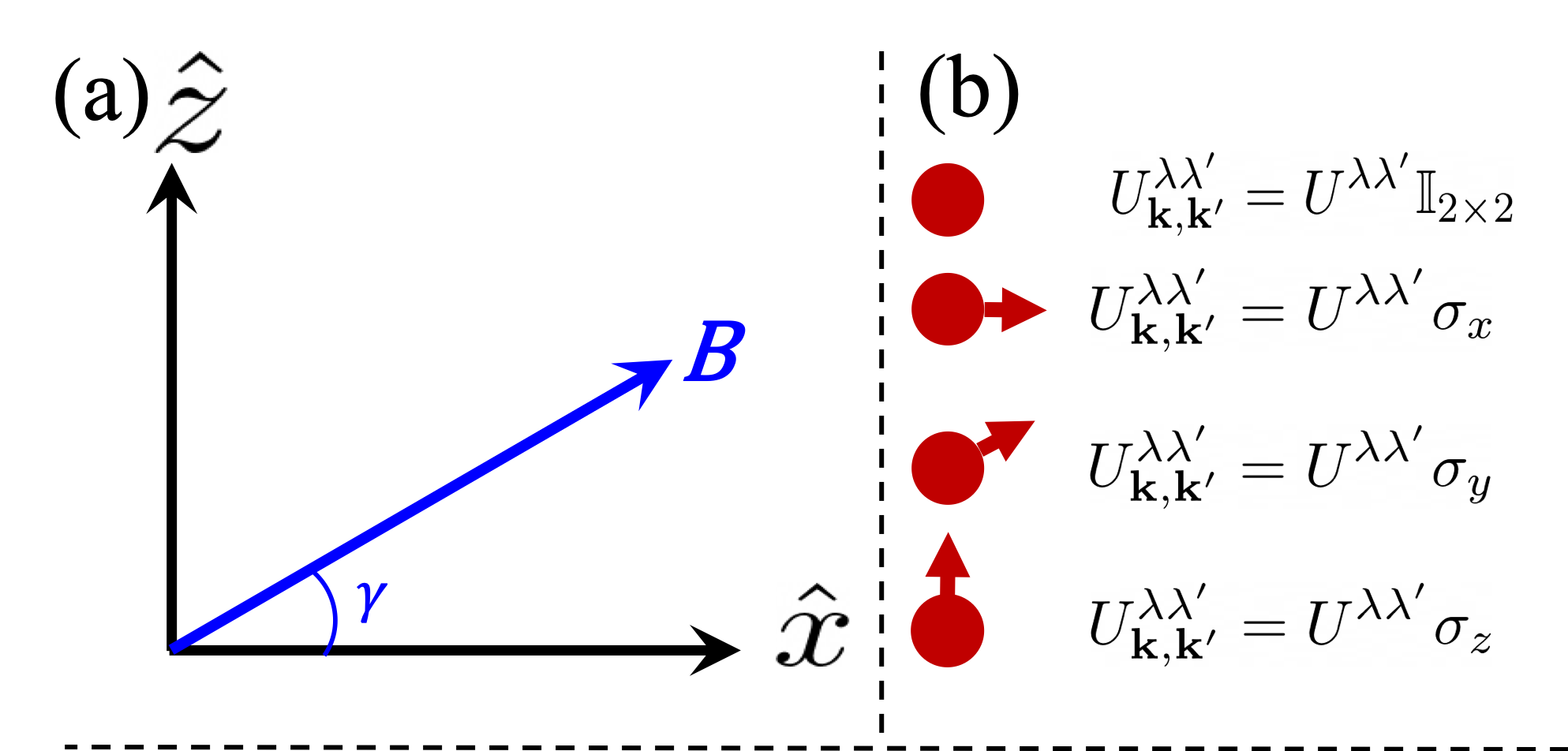}
    \caption {(a) Orientation of magnetic field in the $xz-$plane. (b) Symbolic interpretation for the types of impurity studied in this manuscript (Eq.~\ref{Fermi_gilden_rule}).}
\label{fig:Orientation_magnetic_field}
\end{figure}

\section{Model and formalism}
We begin with the following extended model of a spin-orbit coupled non-centrosymmetric metal that can be expressed near the high-symmetry point as 
\begin{align}
    {H}_\mathrm{soc}(\mathbf{k})=\frac{\hbar^2k^2}{2m}\sigma_{0}+ \hbar\vartheta \mathbf{k}\cdot\mathbf{\sigma} + \hbar \vartheta( k_{x} t_{x} +  k_{z} t_{z})\sigma_{0}
    \label{Hsocncm}
\end{align}
Here, $m$ is the effective electron mass. The second term represents the spin-orbit coupling term, and $\mathbf{\sigma}$ denotes the vector of Pauli matrices in the spin space. The third term in the Hamiltonian tilts the dispersion along a particular direction, and the dimensionless parameters $t_{x}$ and $t_{z}$ represent the tilting along $x$- and $z$-direction, respectively. Similar to the case of WSMs, the tilt term may arise naturally in the bandstructure in SOC-NCMs or may model the effect of strain in the material. 

It is instructive to also compare the above Hamiltonian to a Rashba coupled system given by 
\begin{align}
    H_\mathrm{Rashba}(\mathbf{k}) = \frac{\hbar^2k^2}{2m}\sigma_{0} + \alpha_R (k_x\sigma_y - k_y\sigma_x),
    \label{Hrashba}
\end{align}
where $\alpha_R$ is the Rashba coefficient. The spin-texture for both $H_\mathrm{soc}$ and $H_\mathrm{Rashba}$ can be evaluated as $\mathbf{S}^\lambda=\langle{u^{\lambda}(\mathbf{k})}|\boldsymbol{\sigma}|{u^{\lambda}(\mathbf{k})}\rangle$, where $\ket{u^{\lambda}(\mathbf{k})}$ is the spinor wavefunction and $\lambda$ is the band-index.
The spin-texture for both the above Hamiltonians (Eq.~\ref{Hsocncm} and Eq.~\ref{Hrashba}) is given in Fig.~\ref{fig:spin_text}. The spin rotates as we traverse along the Fermi surface. Importantly, the spins in the two Fermi surfaces point in the opposite direction to each other, indicating their opposite chirality. 

To compare our results with WSM, we use the following prototype model of a two-node time-reversal symmetry broken WSM.
\begin{align}
    {H}_\mathrm{wsm} = \left(\sum\limits_{\chi=\pm 1}\hbar v_F \chi\mathbf{k}\cdot\boldsymbol{\sigma}\right) + \hbar v_F( k_{x} t_{x} +  k_{z} t_{z})\sigma_{0}.
\end{align}
Here, $\chi$ is the chirality and $v_F$ is the Fermi velocity. 
The Hamiltonian in Eq.\ref{Hsocncm} has the following energy dispersion
\begin{align}
    \epsilon^{\lambda} (\mathbf{k})= \frac{\hbar^2k^2}{2m}+\lambda \hbar \vartheta k +  + \hbar \vartheta( k_{x} t_{x} +  k_{z} t_{z})
    \label{KWSM_Hamiltonion}
\end{align}
Here, $\lambda=\mp 1$ is the band index. 
The corresponding eigenvectors are: $\ket{u^{\lambda}}^{T}=[\lambda e^{-i\phi}\cos(\theta /2),\sin(\theta/2)]$. We assume that the Fermi energy $\epsilon_F$ lies above the nodal point $\mathbf{k}=0,$ and thus we have two Fermi surfaces corresponding to the two energy bands as shown in Fig.~\ref{fig:Kramer_weyl_Node_f} (b). The Berry curvature ($\boldsymbol{\Omega}_\mathbf{k}^\lambda$) for both these surfaces has equal magnitudes and opposite signs, just like the Fermi surfaces in the vicinity of two nodal points in WSMs. Interestingly, the orbital magnetic moment ($\mathbf{m}^\lambda_\mathbf{k}$) carries the same sign and magnitude, distinct from WSMs where the signs are reversed. In the presence of an external magnetic field ($\mathbf{B}$), the orbital magnetic moment couples to the dispersion as $-\mathbf{m}^\lambda_\mathbf{k}\cdot\mathbf{B}$ leading to the oval-shaped Fermi surfaces as shown in Fig.~\ref{fig:Kramer_weyl_Node_f} (b). In WSMs, the coupling is opposite, and thus, the shapes of the surfaces are reversed.

We study charge transport in the presence of perturbative electric and magnetic fields using the quasiclassical Boltzmann formalism. This is valid in the limits of weak magnetic fields, $B\ll B_c$, where $eB_c\hbar/2m \epsilon_F=1$.
The non-equilibrium distribution function $f^{\lambda}_{\mathbf{k}}$ obeys the following steady-state equation:
\begin{align}
{\Dot{\mathbf{r}}^\lambda_{\mathbf{k}}}\cdot \mathbf{\nabla_r}{f^{\lambda}_{\mathbf{k}}}+\Dot{\mathbf{k}}^\lambda\cdot \mathbf{\nabla_k}{f^{\lambda}_{\mathbf{k}}}=I_\mathrm{coll}[f^{\lambda}_{\mathbf{k}}].
\label{MB_equation}
\end{align} 
Here, $f^{\lambda}_\mathbf{k} = f_{0} + g^{\lambda}_{\mathbf{k}}$, with $f_{0}$ being the Fermi-Dirac distribution and $g^{\lambda}_{\mathbf{k}}$ is the deviation due to the presence of the external fields. We restrict ourselves to the first order in the electric field, i.e., 
$ g^{\lambda}_{\mathbf{k}} = -e\left({\dfrac{\partial f_{0}}{\partial {\mu}}}\right)_{\epsilon_F}\mathbf{E}\cdot \mathbf{\Lambda^{\lambda}_k}$.
The collision integral ($I_\mathrm{coll}$) in Eq.~\ref{MB_equation} is chosen in such a way that it can incorporate both interband and intraband scattering, given by
\begin{align}
 I_\mathrm{coll}[f^{\lambda}_{\mathbf{k}}]=\sum_{\lambda'} \sum_{\mathbf{k}'}{\mathbf{W}^{\lambda \lambda'}_{\mathbf{k k'}}}{(f^{\lambda'}_{\mathbf{k'}}-f^{\lambda}_{\mathbf{k}})},
 \label{Collision_integral}
\end{align}
where, the scattering rate ${\mathbf{W}^{\lambda \lambda'}_{\mathbf{k k'}}}$ calculated using the Fermi's golden rule,
\begin{align}
{\mathbf{W}^{\lambda \lambda'}_{\mathbf{k k'}}} = \frac{2\pi n}{\mathcal{V}}|\bra{u^{\lambda'}(\mathbf{k'})}U^{\lambda \lambda'}_{\mathbf{k k'}}\ket{u^{\lambda}(\mathbf{k})}|^2\delta(\epsilon^{\lambda'}_{\mathbf{k'}}-\epsilon_F)
\label{Fermi_gilden_rule}
\end{align}
Here, \lq n\rq~is the impurity concentration, \lq $\mathcal{V}$\rq~is the system volume, $\ket{u^{\lambda}(\mathbf{k})}$ is the spinor wavefunction, $U^{\lambda \lambda'}_{\mathbf{k k'}}$ is the scattering potential profile, and $\epsilon_F$ is the Fermi energy. Here we choose $U^{\lambda \lambda'}_{\mathbf{k k'}}$ in such a manner that it can include both magnetic and non-magnetic point-like scattering centers. In general $U^{\lambda \lambda'}_{\mathbf{k k'}} = U^{\lambda \lambda'} \sigma_{i}$ with $i=0,1,2,3$, where $U^{\lambda \lambda'}$ distinguishes the interband ($\lambda\neq\lambda'$) and intraband ($\lambda=\lambda'$) scattering. Here, we work in the geometry represented in Fig.~\ref{fig:Orientation_magnetic_field} (a), i.e., we fix the direction of the electric field along the $z-$direction and rotate the magnetic field in the $xz-$plane that makes an angle $\gamma$ with respect to the $x-$axis. 
Further calculation details for the solution of the distribution function $f^{\lambda}_{\mathbf{k}}$ are presented in the Appendix. Finally, the current is evaluated as $\mathbf{j} = -e\sum\limits_\chi\sum\limits_\mathbf{k} \dot{\mathbf{r}}^\chi f^\chi_\mathbf{k}$, and the conductance tensor $\hat{\sigma}$ is given by $j_\alpha = \sigma_{\alpha\beta} E_\beta$. Unless otherwise specified, we choose the following values for our calculations: $m=10^{-32}\mathrm{kg}$, $\vartheta=5\times 10^{5}\mathrm{ms}^{-1}$, $v_F=10^{6}\mathrm{ms}^{-1}$, $\epsilon_\mathrm{F}=50 \mathrm{meV}$.

\section{Results} 
\subsection{Longitudinal Magnetoconductance}
\begin{figure}
    \centering
    \includegraphics[width=.98\columnwidth]{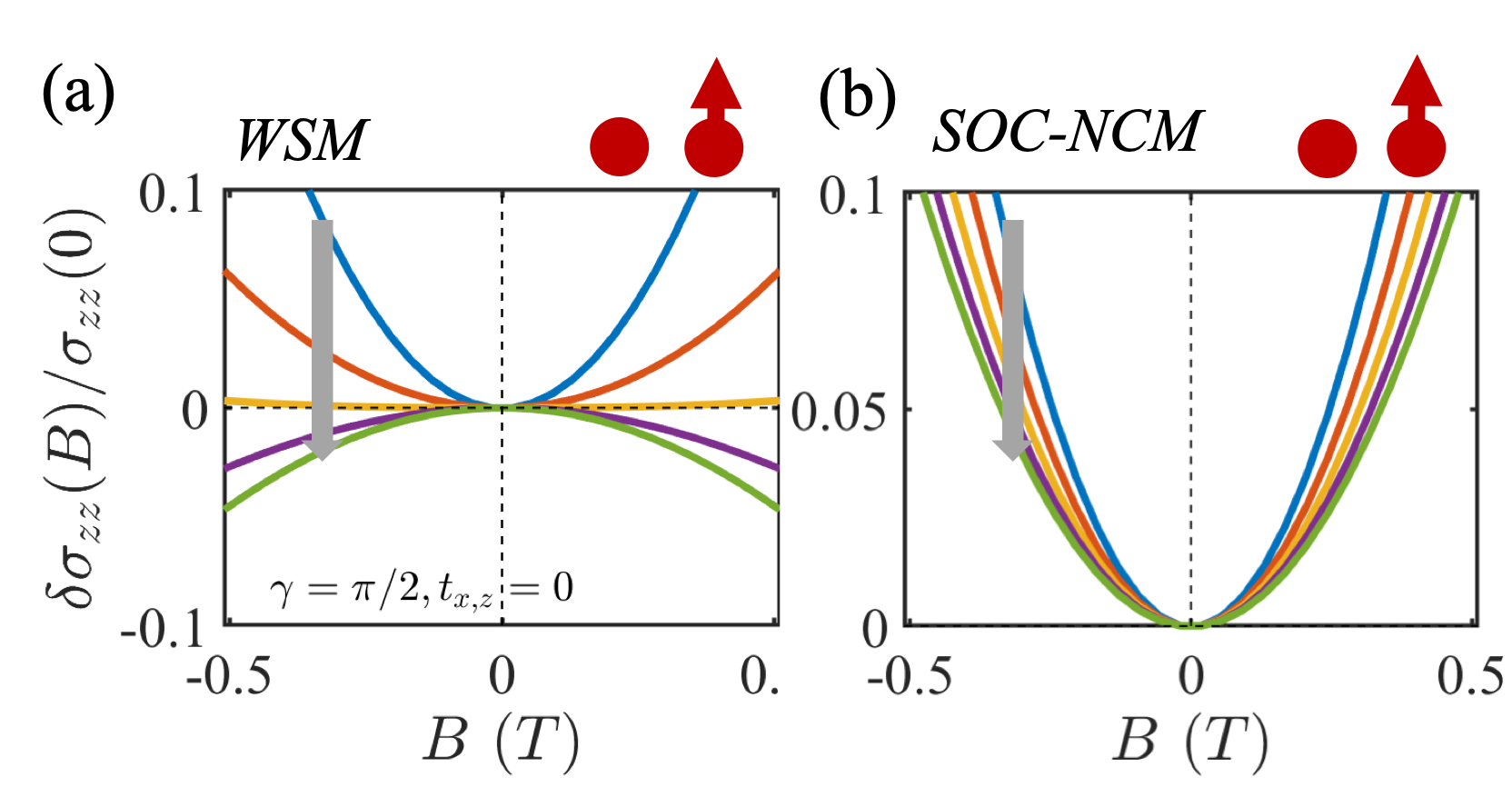}
    \caption{LMC in Weyl semimetals and SOC-NCMs as a function of the magnetic field for different values of the relative interband (internode for WSMs) scattering strengths $\alpha$.  As we move in the direction of the arrow from the blue to the green curve, $\alpha$ is increased from $0.35$ to $1.25$. We obtain the same behavior for a non-magnetic impurity profile i.e., $U_{\mathbf{k,k'}}^{\lambda\lambda'} = U^{\lambda\lambda'} \sigma_0$, as well as magnetic impurity  i.e, $U_{\mathbf{k,k'}}^{\lambda\lambda'} = U^{\lambda\lambda'} \sigma_{z}$. (a) For WSM there is strong-sign-reversal above $\alpha  > \alpha_{c}$, (b) For SOC-NCM, there is no sign-reversal for any interband scattering strength.}
\label{fig:WSM_KWSM_LMC_vs_B_alp_vary_gm_piby2}
\end{figure}

We first discuss longitudinal magnetoconductance for SOC-NCM and compare it with a standard WSM. We examine the behavior of each impurity type (magnetic and non-magnetic) individually. In Fig.~\ref{fig:WSM_KWSM_LMC_vs_B_alp_vary_gm_piby2} (b) we plot the LMC in SOC-NCM as a function of the magnetic field for different values of the relative interband scattering strengths $\alpha$ (the ratio of interband scattering strength to intraband scattetring strength), for non-magnetic $\sigma_0-$impurities as well as $\sigma_z-$magnetic impurities. The LMC is always positive for any value of $\alpha$. This is in striking contrast to WSMs where LMC changes sign (strong sign-reversal~\cite{ahmad2023longitudinal}) when $\alpha>\alpha_c$ (Fig.~\ref{fig:WSM_KWSM_LMC_vs_B_alp_vary_gm_piby2} (a)).

In WSMs, the sign of the OMM is different at Fermi surfaces at both nodes. This breaks the symmetry between them, thus also breaking the symmetry between their chiral partners. This has been attributed to result in a strong-sign-reversal of LMC~\cite{knoll2020negative,sharma2020sign,ahmad2023longitudinal}. 
On the other hand, OMM shifts the energy dispersion in SOC-NMCs in both the Fermi surfaces by the same amount as shown in Fig.~\ref{fig:Kramer_weyl_Node_f} (b). The symmetry between the chiral partners between both the Fermi surfaces thus remains intact and therefore no sign-reversal is observed. Since the magnetic$-\sigma_z$ impurity doesn't flip the chirality of the quasiparticles in both SOC-NCMs and WSMs, we observe the same effect on LMC as a nonmagnetic impurity ($\sigma_0$). The $\sigma_x$ and $\sigma_y$ impurities, on the other hand, flip the chirality of the quasiparticles. We obtain quadratic and positive LMC for both WSMs and SOC-NCMs for $\sigma_x$ and $\sigma_y$ impurities (not plotted explicitly). 

\begin{figure}
    \centering
    \includegraphics[width=.98\columnwidth]{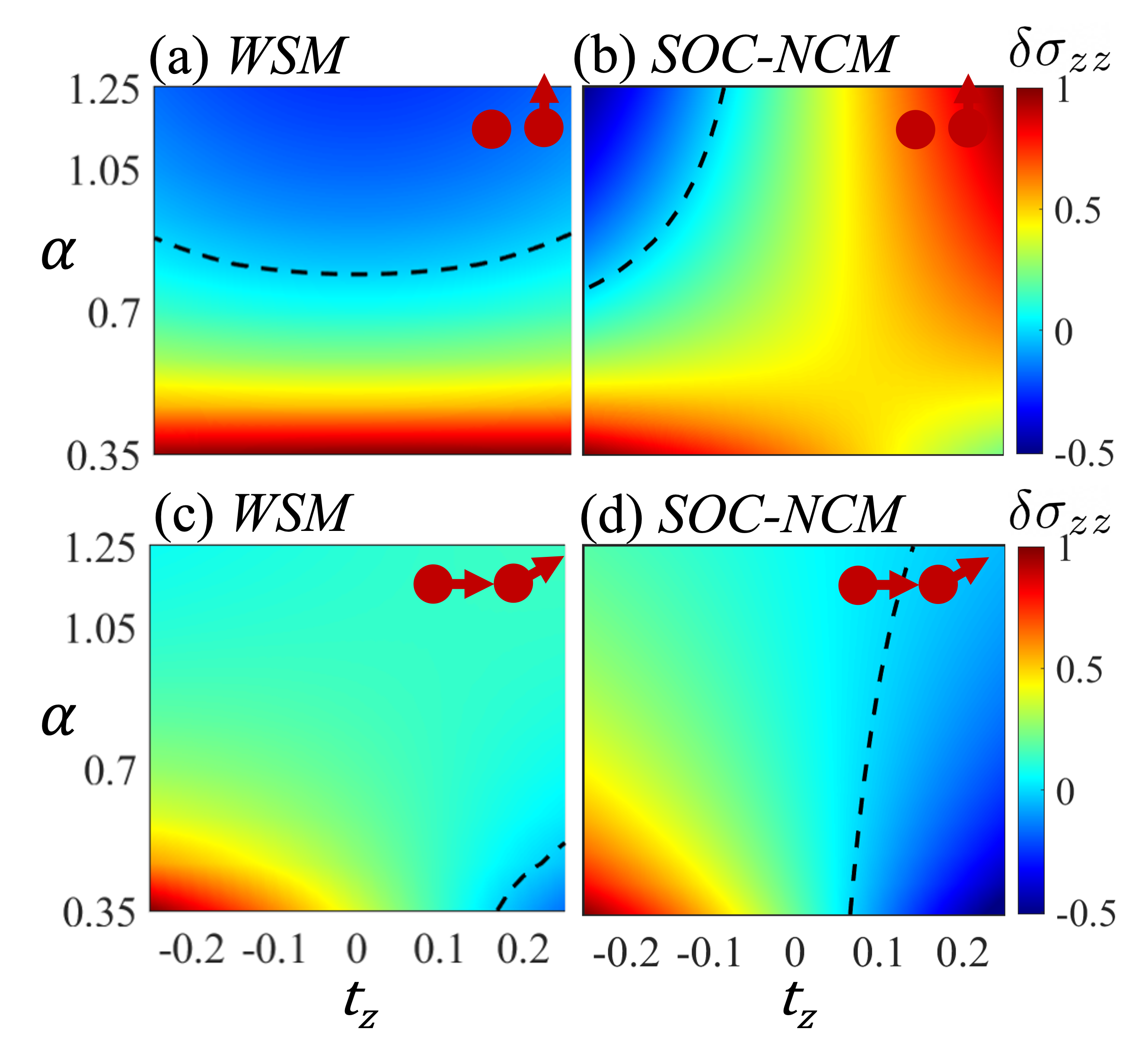}
    \caption{LMC in Weyl semimetals and SOC-NCMs as a function of the relative interband (internode for WSMs) scattering strengths $\alpha$ and the parameter $t_z$.  The dashed black line shows the contour separating positive and negative LMC regions. }
\label{fig:WSM_KWSM_LMC_vs_alpha_vs_tz}
\end{figure}

Next, we examine LMC as a function of the parameter $t_z$ and $\alpha$ for $\sigma_0$ and $\sigma_z$ impurites in Fig.~\ref{fig:WSM_KWSM_LMC_vs_alpha_vs_tz}. We obtain strikingly different behavior for WSMs and SOC-NCMs. For WSMs, the zero-LMC contour $\alpha_c(t_z)$, which separates positive and negative LMC, is now a function of $t_z$ (Fig.~\ref{fig:WSM_KWSM_LMC_vs_alpha_vs_tz} (a)). This change of sign corresponds to strong-sign-reversal.  For SOC-NCMs, the zero-LMC contour appears for nonzero values of $t_z$ and this change of sign is associated with weak-sign-reversal (Fig.~\ref{fig:WSM_KWSM_LMC_vs_alpha_vs_tz} (b)). Specifically, the $t_z$-term tilts the parabola along a particular direction but does not flip its orientation. 

In Fig.~\ref{fig:WSM_KWSM_LMC_vs_alpha_vs_tz} (c) and Fig.~\ref{fig:WSM_KWSM_LMC_vs_alpha_vs_tz} (d) we examine the behavior of LMC as a function of $\alpha$ and $t_z$ for $\sigma_x$ and $\sigma_y-$magnetic impurities. In WSMs, we observe weak-sign-reversal for large values of the tilt parameter $t_z$, and no sign-reversal for smaller values of $t_z$. Furthermore, increasing internode scattering restores positive LMC. This is in sharp contrast to the effect of $\sigma_0$ and $\sigma_z$ impurities in WSMs, where we observe strong-sign-reversal and decreasing internode scattering restoring positive LMC.
In SOC-NCMs, for $\sigma_x$ and $\sigma_y$ impurities, the effect of weak-sign-reversal is more pronounced as shown in Fig.~\ref{fig:WSM_KWSM_LMC_vs_alpha_vs_tz} (d). Again, larger interband scattering restores positive LMC. This feature is understood as follows. The $\sigma_x$ (or $\sigma_y$) impurities flip the chirality of the fermions; further imposing interband scattering back-flips the reversed chirality, and thus interband $\sigma_x$ scattering behaves like intraband scattering.

\begin{figure}
    \centering
    \includegraphics[width=\columnwidth]{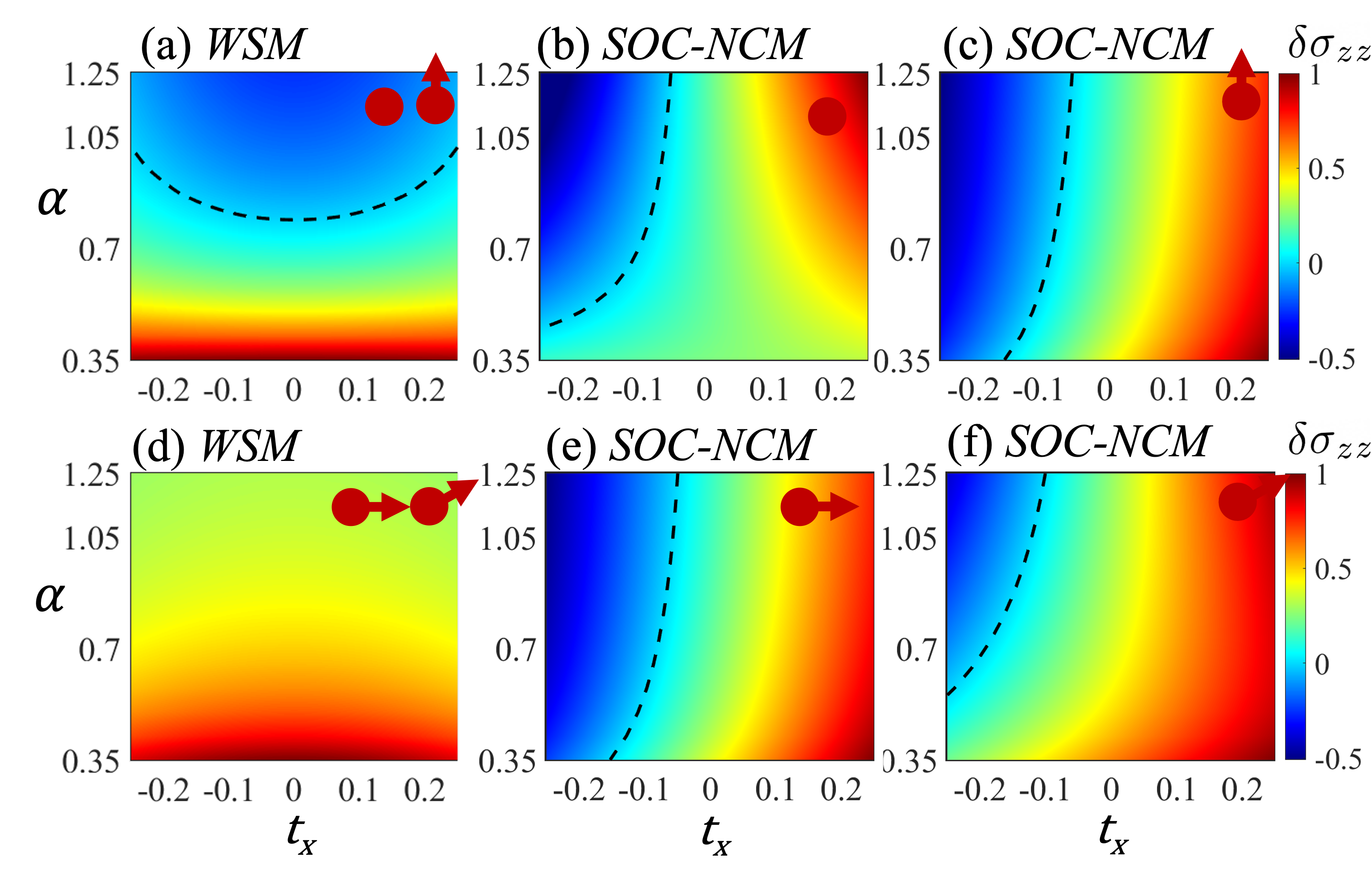}
    \caption{LMC in WSMs and SOC-NCMs in the presence of $t_x$ parameter. The dashed black line shows the contour separating positive and negative LMC regions.}
    \label{fig:lmc_tx_alpha_color}
\end{figure}
\begin{figure*}
    \centering
    \includegraphics[width=1.99\columnwidth]{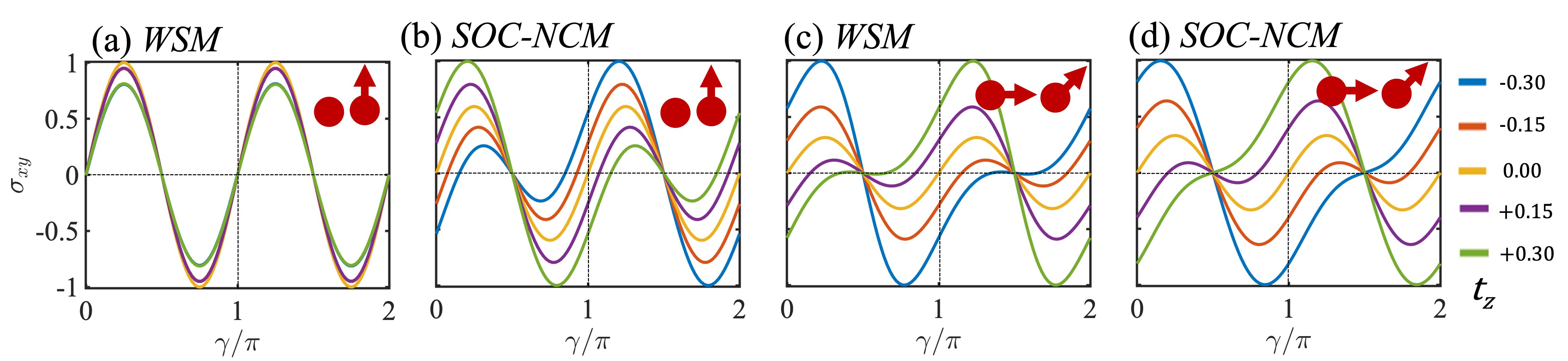}
    \caption{Planar Hall conductivity for WSM and SOC-NCM for different impurity types in the presence of parameter $t_z$. Plots are appropriately normalized.}
    \label{fig:phc_01}
\end{figure*}
\begin{figure}
    \centering
    \includegraphics[width=\columnwidth]{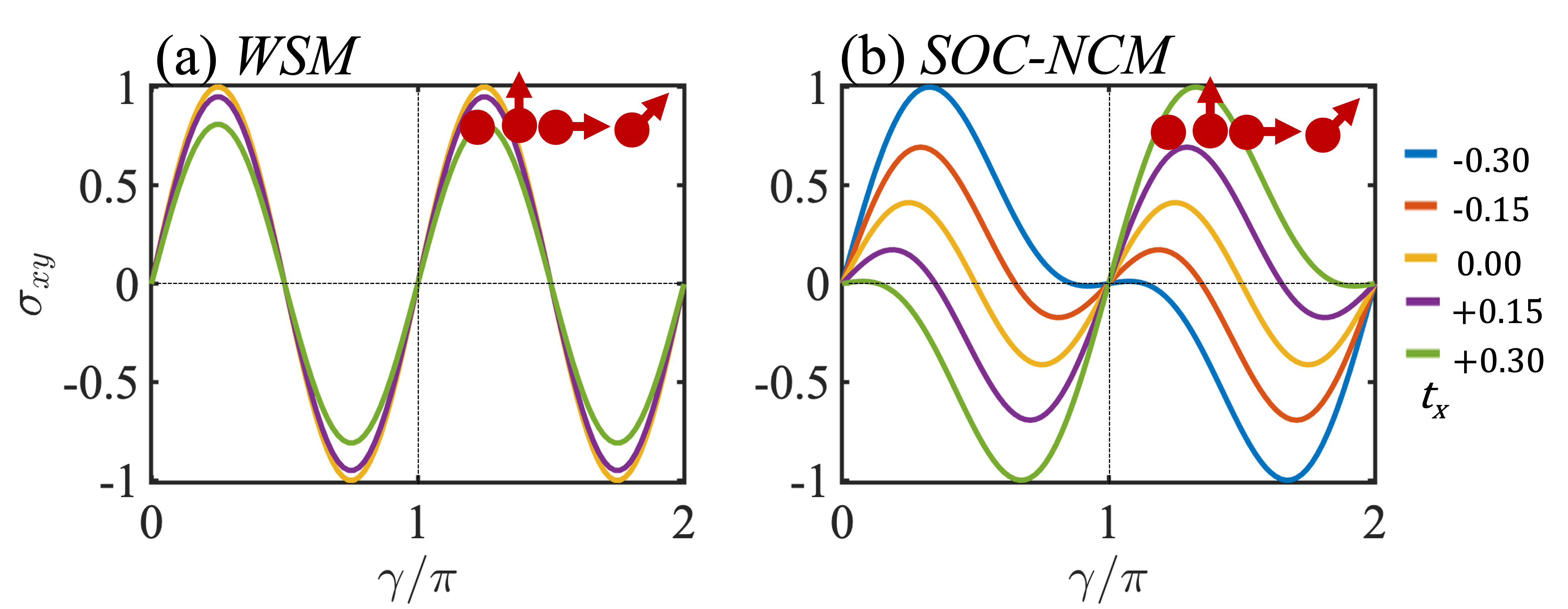}
    \caption{Planar Hall conductivity for WSM and SOC-NCM for different impurity types in the presence of parameter $t_x$. Plots are appropriately normalized.}
    \label{fig:phc_02}
\end{figure}

In Fig.~\ref{fig:lmc_tx_alpha_color} we compare the behavior of LMC of WSM with that of SOC-NCM in the presence of the tilt parameter $t_x$. In WSM, just like the case when $t_z\neq 0$, we find that the behavior in the presence of non-magnetic and $\sigma_z-$impurities is similar; we observe strong-sign-reversal. The zero-LMC contour $\alpha_c(t_x)$ is qualitatively similar to the case of nonzero $t_z$, but nevertheless exhibits quantitative differences. In SOC-NCM, weak-sign-reversal is observed. Unlike the $t_z\neq 0$ case, we observe quantitative differences between non-magnetic and $\sigma_z-$impurities (Fig.~\ref{fig:lmc_tx_alpha_color} (b) and Fig.~\ref{fig:lmc_tx_alpha_color} (c)). Surprisingly, WSMs in the presence of $\sigma_x$ or $\sigma_y$ impurities, exhibit neither weak-sign-reversal nor strong-sign-reversal in the presence of $t_x$ parameter (Fig.~\ref{fig:lmc_tx_alpha_color} (d)). This can be again understood from the fact that either of these impurity types changes the roles of internode scattering and that tilting the Weyl cone in a direction orthogonal to the direction of the magnetic field doesn't add an overall linear component to the magnetoconductivity. For SOC-NCM, we observe weak-sign-reversal, with quantitative differences between $\sigma_x$ and $\sigma_y$ impurities.

\subsection{Planar Hall conductance}
We next discuss the planar Hall conductance in SOC-NCM and compare the results with a standard WSM. In WSMs, the PHC can be expressed as~\cite{ahmad2023longitudinal}
\begin{align}
\sigma_{xz}(B)= \sigma_{xz}^{(2)}(B-B_0)^2 + \sigma_{xz}^{(0)},
\label{eq:PHC1}
\end{align} 
where $B_0$ is the vertex of the parabola, and $\sigma_{xz}^{(2)}$ is the quadratic coefficient. The above form allows us to generalize PHC away from the origin, i.e., $B_0\neq0$. The angular dependence for WSM is $\sin{(2\gamma)}$ for a point-like non-magnetic impurity profile. We find that this dependence is retained for magnetic impurities pointing in the $z-$direction as well (Fig.~\ref{fig:phc_01} (a)), and tilting the Weyl cones ($t_z\neq 0$) only has a quantitative effect. In contrast, SOC-NCMs have a qualitatively different dependence on $t_z$ (Fig.~\ref{fig:phc_01} (b)). We observe that unlike in WSMs, the planar Hall conductance in SOC-NCMs exhibits weak-sign-reversal as a function of the parameter $t_z$. For $\sigma_x$ and $\sigma_y$ impurities, PHC in both WSMs and SOC-NCMs exhibit weak-sign-reversal and exhibit similar qualitative behavior (Fig.~\ref{fig:phc_01} (c) and (d)). In both systems, interband (internode) scattering is found to have no significant qualitative effect on planar Hall conductance. 

For the nodes tilted along the $x-$direction, we observe qualitatively very different behavior. For $\sigma_0$ point-like impurities, the $\sin 2\gamma$ trend is observed irrespective of the value of $t_x$, as expected for Weyl cones that are oriented in the same direction. We find qualitatively similar behavior irrespective of the impurity type (Fig.~\ref{fig:phc_02} (a)). Note that if the Weyl cones were oriented opposite to each other, one instead finds a $\sin \gamma$ behavior of PHC~\cite{nandy2017chiral}. In the case of SOC-NCMs, one finds a transition from $\sin 2\gamma$ trend to $\sin \gamma$ as the parameter $t_x$ is increased from zero in either direction (Fig.~\ref{fig:phc_02}). Like WSMs, we observe qualitatively similar behavior for both magnetic (any direction) and non-magnetic impurities. 
\begin{figure}
    \centering
    \includegraphics[width=0.9\columnwidth]{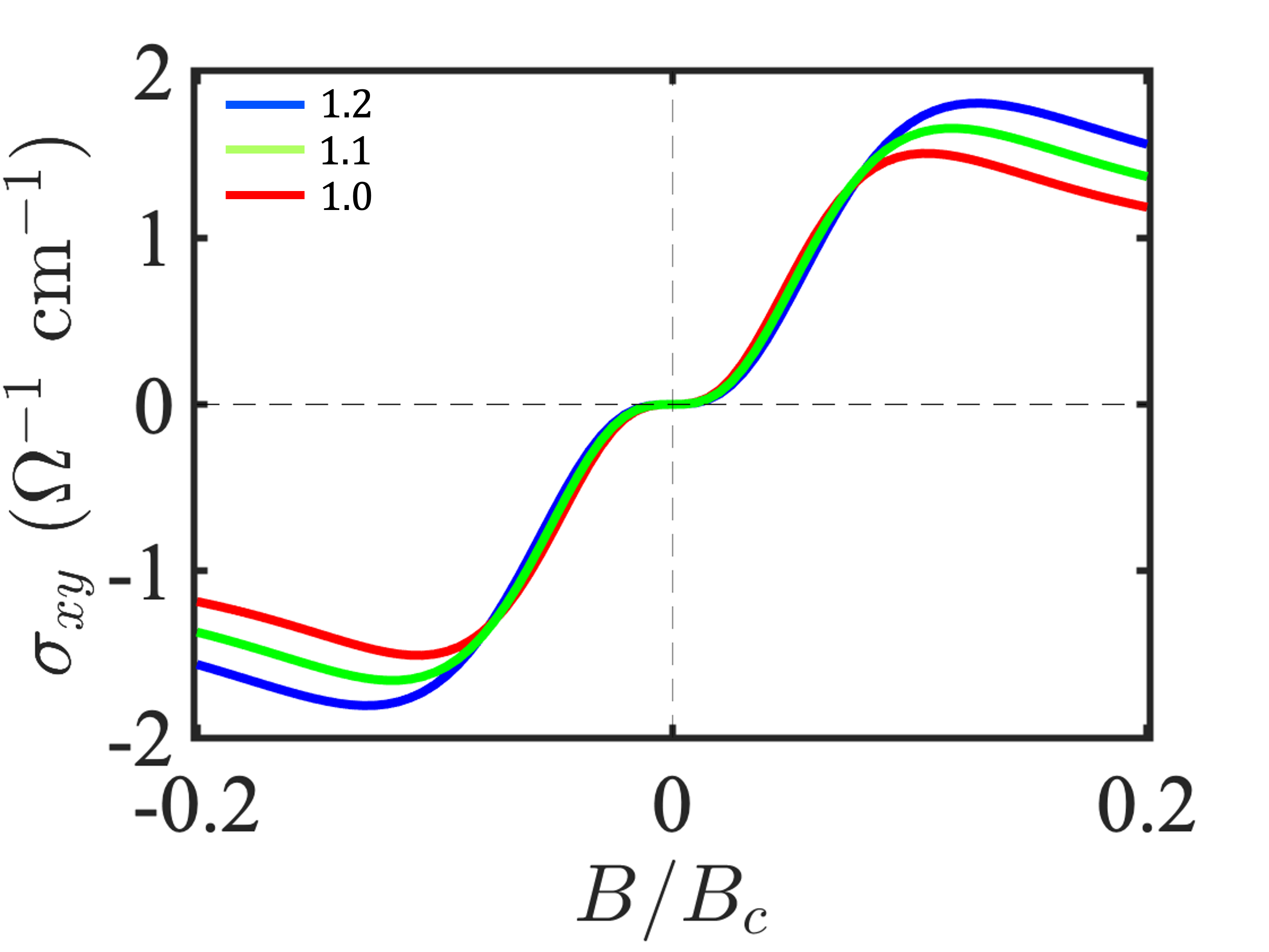}
    \caption{Geometrical contribution to the Hall conductivity showing a non-monotonic behavior. The legends indicate the strength of spin-orbit coupling parameter $\vartheta$ in units of $10^5 \mathrm{ms}^{-1}$.  We chose $g=2$.}
    \label{fig:ahe_01}
\end{figure}
\subsection{Anomalous contribution to the Hall Conductance}
In WSMs, the nonvanishing anomalous Hall conductance (AHC) has been attributed to the presence of a finite vector $\mathbf{k}_0$ that separates Weyl cones of opposite chiralities. The net AHC is given by $\sigma_{xy}^a = e^2 k_0/\hbar$. In the presence of time-reversal symmetry, multiple such vectors add up to zero, and AHC is zero.  It is noteworthy that the intrinsic AHC contribution of one node, which is given by the integral of the Berry curvature of the filled band up to the Fermi surface, exactly cancels the contribution of the other node. The nonzero AHC in TR-broken WSMs is understood by considering a gapped 2D Chern insulator $H(\mathbf{k}_\perp,k_z)$ that undergoes a topological phase transition at the Weyl node~\cite{wan2011topological}. 

In SOC-NCMs, we particularly focus on a single nodal point. It is expected that the two Fermi surfaces that enclose the nodal point cancel out their contributions of anomalous Hall conductivity, but as we show next, the presence of an external magnetic field induces finite anomalous contribution to the Hall conductivity.
In the presence of an external magnetic field, Zeeman coupling will introduce an additional term in the Hamiltonian given by~\cite{sharma2017nernst,sharma2018tunable,gadge2022anomalous}
\begin{align}
    H_z = -g\mu_B\boldsymbol{\sigma}\cdot\mathbf{B}.
    \label{Hz}
\end{align}
This causes an opposite energy shift in both bands. 
Furthermore, the anomalous shift in the energy dispersion due to the orbital magnetic moment ($\epsilon^\lambda_\mathbf{k}\rightarrow\epsilon^\lambda_\mathbf{k}-\mathbf{m}^\lambda_\mathbf{k}\cdot \mathbf{B}$). Both of these effects, in concurrence, lead to a finite and measurable anomalous contribution to the Hall conductance, which at zero temperature is calculated as 
\begin{align}
    \sigma_{xy}^a = \frac{e^2}{\hbar} \sum\limits_{\lambda=\pm 1}\int{\frac{d^3\mathbf{k}}{(2\pi)^3}} \mathcal{D}^\lambda_\mathbf{k}\theta(\epsilon_F - \epsilon^\lambda_\mathbf{k}) \Omega^\lambda_z(\mathbf{k}),
    \label{Eq_ahe}
\end{align}
where $\mathcal{D}^\lambda_\mathbf{k}=(1+e\mathbf{B}\cdot\boldsymbol{\Omega}^\lambda_\mathbf{k}/\hbar)$. 
In Fig.~\ref{fig:ahe_01} we plot the geometrical contribution to the Hall conductivity for SOC-NCM as evaluated from Eq.~\ref{Eq_ahe}. Interestingly, the behavior is non-monotonic with respect to the magnetic field, and for low enough magnetic field, the anomalous contribution is found to be independent of the strength of the spin-orbit coupling parameter. The non-monotinicity can be understood as follows. With increasing magnitude of the magnetic field, the net Berry curvature contribution increases, but eventually decreases as for larger magnetic fields as the magnitude of the Berry curvature itself reduces. This can be easily tested in current and upcoming transport experiments in SOC-NCMs.

\section{Conclusions and Discussions}
Chiral anomaly is a Fermi surface property with similar characteristics in Weyl and spin-orbit coupled noncentrosymmetric metals. It manifests itself in the measurement of longitudinal magnetoconductance and the planar Hall conductance. 
However, in striking contrast to WSMs, where the sign of the LMC is sensitive to the internode scattering strength, the sign of the LMC in SOC-NCMs is independent of the interband scattering strength and always remains positive. The reason is traced down to the subtle difference in the orbital magnetic moment in WSMs and SOC-NCMs. Orbital magnetic moment in SOC-NCMs is of equal magnitudes and signs in both the bands but has opposite signs at the two nodes in WSMs. This difference also yields drastic differences in other transport properties, such as the planar Hall conductivity. We also examined all the properties in the presence of a tilt parameter ($t_z$ and $t_x$) and for different impurity types (magnetic and non-magnetic). The behavior for $\sigma_0$ and $\sigma_z$ impurities was found to be qualitatively similar to each other as they do not flip with chirality. On the other hand, $\sigma_x$ and $\sigma_y$ flip the chirality and behave qualitatively similar to each other. Lastly, we predict that the combination of the anomalous orbital magnetic moment and the Zeeman field gives rise to a geometrical contribution to the Hall conductivity in SOC-NCMs that is non-monotonic in the magnetic field. Our study is highly pertinent in light of current and upcoming experiments in the field of spin-orbit coupled noncentrosymmetric and Weyl metals.

\textit{Acknowledgements}
G.V.K. and A.A. acknowledge financial support from IIT Mandi HTRA. G.S.
acknowledges support from IIT Mandi Seed Grant. We thank Shubhanshu Karoliya for his technical support.  

\appendix
\section{Maxwell Boltzmann transport theory}
Due to Berry phase effects, in the presence of electric and magnetic fields, the semiclassical dynamics of the Bloch electrons are modified and governed by the following equation~\cite{son2012berry,knoll2020negative}.
\begin{align}
\dot{\mathbf{r}}^\lambda &= \mathcal{D}^\lambda \left( \frac{e}{\hbar}\left(\mathbf{E}\times \boldsymbol{\Omega}^\lambda \right)+ \frac{e}{\hbar}(\mathbf{v}^\lambda\cdot \boldsymbol{\Omega}^\lambda) \mathbf{B} + \mathbf{v}_\mathbf{k}^\lambda\right) \nonumber\\
\dot{\mathbf{p}}^\lambda &= -e \mathcal{D}^\lambda \left( \mathbf{E} + \mathbf{v}_\mathbf{k}^\lambda \times \mathbf{B} + \frac{e}{\hbar} \left(\mathbf{E}\cdot\mathbf{B}\right) \boldsymbol{\Omega}^\lambda \right).
\label{Couplled_equation}
\end{align}
where $\mathbf{v}_\mathbf{k}^\lambda = \frac{1}{\hbar}\frac{\partial\epsilon^{\lambda}(\mathbf{k})}{\partial\mathbf{k}}$ is the band velocity, $\boldsymbol{\Omega}^\lambda = -\lambda \mathbf{k} /2k^3$ is the Berry curvature, and $\mathcal{D}^\lambda = (1+e\mathbf{B}\cdot\boldsymbol{\Omega}^\lambda/\hbar)^{-1}$ is the factor modifying the density of the states in the presence of the Berry curvature. The self-rotation of the Bloch wave packet also gives rise to an orbital magnetic moment, $\mathbf{m}^\lambda_\mathbf{k}$~\cite{xiao2010berry}. In the presence of a magnetic field, the orbital magnetic moment shifts the energy dispersion as $\epsilon^{\lambda}_{\mathbf{k}}\rightarrow \epsilon^{\lambda}_{\mathbf{k}} - \mathbf{m}^\lambda_\mathbf{k}\cdot \mathbf{B}$. 
Using Eq.~\ref{Couplled_equation} and Eq.~\ref{Collision_integral} and retaining terms only up to linear order in electric and magnetic fields, the Boltzmann transport equation becomes
\begin{align}
&\left[\left(\frac{\partial f_0^\lambda}{\partial \epsilon^\lambda_\mathbf{k}}\right) \mathbf{E}\cdot \left(\mathbf{v}^\lambda_\mathbf{k} + \frac{e\mathbf{B}}{\hbar} (\boldsymbol{\Omega}^\lambda\cdot \mathbf{v}^\lambda_\mathbf{k}) \right)\right]\nonumber\\
 &= -\frac{1}{e \mathcal{D}^\lambda}\sum\limits_{\lambda'}\sum\limits_{\mathbf{k}'} W^{\lambda\lambda'}_{\mathbf{k}\mathbf{k}'} (g^\lambda_{\mathbf{k}'} - g^\lambda_\mathbf{k})
 \label{Eq_boltz2}
\end{align}
We have fixed the direction of the electric field along increasing $x$-direction, and the magnetic field is rotated in $xz$-plane (See Fig.\ref{fig:Orientation_magnetic_field}). Therefore, $\mathbf{E} = E(0,0,1)$ and  $\mathbf{B} = B (\cos{\gamma},0,\sin{\gamma})$. In this case, only the $z$-component of $\mathbf{\Lambda}$ is relevant. Therefore Eq.~\ref{Eq_boltz2} reduces to,
\begin{align}
\mathcal{D}^{\lambda}(k)\left[{v^{\lambda,z}_{\mathbf{k}}}+\frac{eB\sin{\gamma}}{\hbar}(\mathbf{v^{\lambda}_k}\cdot\mathbf{\Omega}^{\lambda}_k)\right]
= \sum_{\lambda' \mathbf{k}'}{\mathbf{W}^{\lambda \lambda'}_{\mathbf{k k'}}}{(\Lambda^{\lambda'}_{\mathbf{k'}}-\Lambda^{\lambda}_{\mathbf{k}})}.
\label{boltzman_in_terms_lambda}  
\end{align}

We define the valley scattering time ($\tau^\lambda_\mathbf{k}$) as follows
\begin{align}
\frac{1}{\tau^{\lambda}_{\mathbf{k}}(\theta,\phi)}=\sum_{\lambda'}\mathcal{V}\int\frac{d^3\mathbf{k'}}{(2\pi)^3}(\mathcal{D}^{\lambda'}_{\mathbf{k}'})^{-1}\mathbf{W}^{\lambda \lambda'}_{\mathbf{k k'}}
\label{Tau_invers}
\end{align}
$\mathbf{W}^{\lambda \lambda'}_{\mathbf{k k'}}$ is defined in Eq.~\ref{Fermi_gilden_rule} and the corresponding overlap of the Bloch wave-function is
$\mathcal{G}_{i}^{\lambda\lambda'}(\theta,\phi) = [1+\lambda\lambda'\xi_{i}(\cos{\theta}\cos{\theta'} + \alpha_{i}\sin{\theta}\sin{\theta'}\cos{\phi}\cos{\phi'} + \beta_{i}\sin{\theta}\sin{\theta'}\sin{\phi}\sin{\phi'}]$ with $i=0,1,2,3$ (Please see Tab.~\ref{tab:Table1}).
\begin{table}[b]
\begin{ruledtabular}
\begin{tabular}{cccccccc}
 i &$\sigma_{i}$ &$\alpha_{i}$ &$\beta_{i}$  &&$\xi_{i}$\\
 \hline
 0 &$\mathbb{I}_{2\times2}$&+1 &+1 && +1\\
 1 &$\sigma_{x}$&-1 &+1 && -1\\
 2 &$\sigma_{y}$&+1 &-1 && -1\\
 3 &$\sigma_{z}$&-1 &-1 && +1\\
\end{tabular}
\caption{\label{tab:Table1}
The signs of $\alpha, \beta$, and $\xi$ are used in the expression of overlap of the Bloch wave function (see Eq.~\ref{Tau_invers}). $\sigma_{x, y, z}$ are the components of the Pauli spin vector, and $\mathbb{I}_{2\times2}$ is the identity matrix. }
\end{ruledtabular}
\end{table}
Taking Berry phase into account and corresponding change in the density of states, $\sum_{k}\longrightarrow \mathcal{V}\int\frac{d^3\mathbf{k}}{(2\pi)^3}\mathcal{D}^\lambda(k)$, Eq.~\ref{boltzman_in_terms_lambda} becomes
\begin{multline}
h^{\lambda}_{\mu}(\theta,\phi) + \frac{\Lambda^{\lambda}_{\mu,i}(\theta,\phi)}{\tau^{\lambda}_{\mu,i}(\theta,\phi)}\\=\sum_{\lambda'}\mathcal{V}\int\frac{d^3\mathbf{k}'}{(2\pi)^3} \mathcal{D}^{\lambda'}(k')\mathbf{W}^{\lambda \lambda'}_{\mathbf{k k'}}\Lambda^{\lambda'}_{\mu,i}(\theta',\phi')
\label{MB_in_term_Wkk'}
\end{multline}
Here $h^{\lambda}_{\mu}(\theta,\phi)=\mathcal{D}^{\lambda_{\mathbf{k}}}[v^{\lambda}_{z,\mathbf{k}}+eB\sin{\gamma}(\mathbf{\Omega}^{\lambda}_{k}\cdot \mathbf{v}^{\lambda}_{\mathbf{k}})]$. In the zero-temperature limit, for a constant Fermi energy surface, Eq.~\ref{Tau_invers} and RHS of Eq.~\ref{MB_in_term_Wkk'} is reduced to the integration over $\theta'$ and $\phi'$:
\begin{align}
\frac{1}{\tau^{\lambda}_{\mu,i}(\theta,\phi)} =  \mathcal{V}\sum_{\lambda'} \Pi^{\lambda\lambda'}\iint\frac{(k')^3\sin{\theta'}}{|\mathbf{v}^{\lambda'}_{k'}\cdot{\mathbf{k'}^{\lambda'}}|}d\theta'd\phi' \mathcal{G}_{i}^{\lambda\lambda'}(D^{\lambda'}_{\mathbf{k'}})^{-1}
\label{Tau_inv_int_thet_phi}
\end{align}
\begin{multline}
\mathcal{V}\sum_{\lambda'} \Pi^{\lambda\lambda'}\iint f^{\lambda'}(\theta',\phi')\mathcal{G}_{i}^{\lambda\lambda'} d\theta'd\phi'\times[d^{\lambda'} - h^{\lambda'}_{\mu}(\theta',\phi') \\+ a^{\lambda'} \cos\theta' + b^{\lambda'} \sin\theta' \cos{\phi'} + c^{\lambda'}\sin{\theta'} \cos{\phi'}] 
\end{multline}
where $\Pi^{\lambda \lambda'} = N|U^{\lambda\lambda'}|^2 / 4\pi^2 \hbar^2$, $f^{\lambda} (\theta,\phi)=\frac{(k)^3}{|\mathbf{v}^\lambda_{\mathbf{k}}\cdot \mathbf{k}^{\lambda}|} \sin\theta (\mathcal{D}^\eta_{\mathbf{k}})^{-1} \tau^\lambda_\mu(\theta,\phi)$. Using ansatz $\Lambda^{\lambda}_{\mathbf{k}}=[d^{\lambda}-h^{\lambda}_{k'} + a^{\lambda}\cos{\phi} +b^{\lambda}\sin{\theta}\cos{\phi}+c^{\lambda}\sin{\theta}\sin{\phi}]\tau^{\lambda}_{\mu}(\theta,\phi)$ the above equation is written in the following form:
\begin{multline}
d^{\lambda}+a^{\lambda}\cos{\phi}+b^{\lambda}\sin{\theta}\cos{\phi}+c^{\lambda} \sin{\theta}\sin{\phi}\\
=\sum_{\lambda'}\mathcal{V}\Pi^{\lambda\lambda'}\iint f^{\lambda'}(\theta',\phi')d\theta'd\phi'\\\times[d^{\lambda'}-h^{\lambda'}_{k'}+a^{\lambda'}\cos{\theta'}+b^{\lambda'}\sin{\theta'}\cos{\phi'}+c^{\lambda'} \sin{\theta'}\sin{\phi'}]\\
\label{Boltzman_final}
\end{multline}
When the aforementioned equation is explicitly put out (for each value of $i$), it appears as seven simultaneous equations that must be solved for eight variables. The particle number conservation provides another restriction:
\begin{align}
\sum\limits_{\lambda}\sum\limits_{\mathbf{k}} g^\lambda_\mathbf{k} = 0
\label{Eq_sumgk}
\end{align} 
For the eight unknowns ($d^{\pm 1}, a^{\pm 1}, b^{\pm 1}, c^{\pm 1}$), equations \ref{Boltzman_final} and \ref{Eq_sumgk} are simultaneously solved with Eq.~\ref{Tau_inv_int_thet_phi}. Due to the intricate structure of the equations, all two-dimensional integrals with respect to $\theta'$ and $\phi'$ the simultaneous equations' solution are carried out numerically.

\pagebreak
 
\bibliography{biblio.bib}
\end{document}